\documentclass[conference]{IEEEtran}
\IEEEoverridecommandlockouts
\usepackage{cite}
\usepackage{amsmath,amssymb,amsfonts}
\usepackage{algorithmic}
\usepackage{graphicx}
\usepackage{textcomp}
\usepackage{xcolor}
\usepackage{balance}
\usepackage{float}

\newcounter{MYtempeqncnt}
\newtheorem{lemma}{Lemma}
\newtheorem{remark}{Remark}
\newtheorem{theorem}{Theorem}

\newtheorem{corollary}{Corollary}

\newtheorem{proposition}{Proposition}
\def\BibTeX{{\rm B\kern-.05em{\sc i\kern-.025em b}\kern-.08em
    T\kern-.1667em\lower.7ex\hbox{E}\kern-.125emX}}

\begin{document}

\title{Channel Estimation and Power Scaling Law of Large Reflecting Surface with Non-Ideal Hardware
}

\author{\IEEEauthorblockN{Yiming Liu, Erwu Liu, Rui Wang and Yuanzhe Geng}
\IEEEauthorblockA{\textit{College of Electronics and Information Engineering, Tongji University, Shanghai, China} \\
Emails: 15995086362@163.com, erwu.liu@ieee.org, ruiwang@tongji.edu.cn, yuanzhegeng@tongji.edu.cn}
}

\maketitle

\begin{abstract}
Large reflecting surface (LRS) has emerged as a new solution to improve the energy and spectrum efficiency of wireless communication system. Most existing studies were conducted with an assumption of ideal hardware, and the impact of hardware impairments receives little attention. However, the non-negligible hardware impairments should be taken into consideration when we evaluate the system performance. In this paper, we consider an LRS assisted communication system with hardware impairments, and focus on the channel estimation study and the power scaling law analysis. First, with linear minimum mean square error estimation, we theoretically characterize the relationship between channel estimation performance and impairment level, number of reflecting elements, and pilot power. After that, we analyze the power scaling law and reveal that if the base station (BS) has perfect channel state information, the transmit power of user can be made inversely proportional to the number of BS antennas and the square of the number of reflecting elements with no reduction in performance; If the BS has imperfectly estimated channel state information, to achieve the same performance, the transmit power of user can be made inversely proportional to the square-root of the number of BS antennas and the square of the number of reflecting elements.
\end{abstract}


\section{Introduction}
Due to the explosive growth of mobile data traffic in recent years, we need to enhance the performance of future wireless communication systems. Many related works have shown that multiple-input multiple-output (MIMO) technology can offer improved energy and spectrum efficiency, owing to both array gains and diversity effects, \textit{e.g.}, Ngo, Larsson and Marzetta prove that the power transmitted by the user can be cut inversely proportional to the square-root of the number of base station (BS) antennas with no reduction in performance \cite{6457363}. However, the requirements of high hardware cost and high complexity are still the main hindrances to its implementation. Recently, large reflecting surface (LRS), \textit{a.k.a.}, intelligent reflecting surface (IRS) has emerged as a new solution to improve the energy and spectrum efficiency of wireless communication system, and can be used as a low-cost alternative to massive MIMO system \cite{8796365,8936989,8930608,8888223,8647620}. Prior works demonstrate that the LRS can effectively control the wavefront, \textit{e.g.}, the phase, amplitude, frequency, and even polarization, of the impinging signals without the need of complex decoding, encoding, and radio frequency processing operations. Basar, \textit{et al.}, elaborate on the fundamental differences of this state-of-the-art solution with other technologies, and explain why the use of LRS necessitates to rethink the communication-theoretic models currently employed in wireless networks \cite{8796365}. {\"O}zdogan, Bj{\"o}rnson and Larsson demonstrate that the LRS can act as diffuse scatterers to jointly beamform the signal in a desired direction in \cite{8936989}. They also compare the LRS with the decode-and-forward (DF) relay, and show that the LRS can achieve higher energy efficiency by using many reflecting elements \cite{8888223}. Wu and Zhang analytically show that the LRS with discrete phase shifts achieve the same power gain with that of the LRS with continuous phase shifts \cite{8930608}. They also verify that the LRS is able to drastically enhance the link quality and/or coverage over the conventional setup without the LRS in \cite{8647620}.

It is noted that all the mentioned works study the LRS systems with an assumption of perfect or ideal hardware operations without any impairments. However, both physical transceiver and LRS suffer from hardware impairments which are non-negligible in practice. Bj{\"o}rnson,\textit{ et al.}, prove that hardware impairments greatly limit the performance of channel estimation and bound the channel capacity of massive MIMO system \cite{6362131,6891254}. To reveal the impact of hardware impairments on the LRS system, in this paper, we focus on the study of channel estimation and the power scaling law analysis by taking the hardware impairment into account. 

With the use of linear minimum mean square error (LMMSE) estimator, our analysis shows that the estimation error decreases with the power of pilot signal, but increases with the number of reflecting elements and the level of hardware impairments. Although the hardware impairments of LRS have no effect on the estimation accuracy statistically, the hardware impairments of transceiver limit the estimation performance when signal-to-noise ratio (SNR) goes to infinity. In addition, the estimation error of LRS channel is larger than that of direct channel. All obtained results imply that more accurate estimation methods and more efficient communication protocols are needed in future works. After that, we analyze the power scaling law of user in the cases of perfect and imperfect channel state information. Our obtained results show that if the BS has perfect channel state information, the transmit power of user can be made inversely proportional to the number of BS antennas and the square of the number of reflecting elements with no reduction in performance, and if the BS has imperfect channel state information from channel estimation, the transmit power of user can be made inversely proportional to the square-root of the number of BS antennas and the square of the number of reflecting elements to achieve the same performance. This is encouraging for that we can use more low-cost reflecting elements instead of expensive antennas to achieve higher power scaling.

\section{Communication System Model}
\begin{figure}[htbp]
	\vspace{-1.25 em}
	\centerline{\includegraphics[width=8.7 cm]{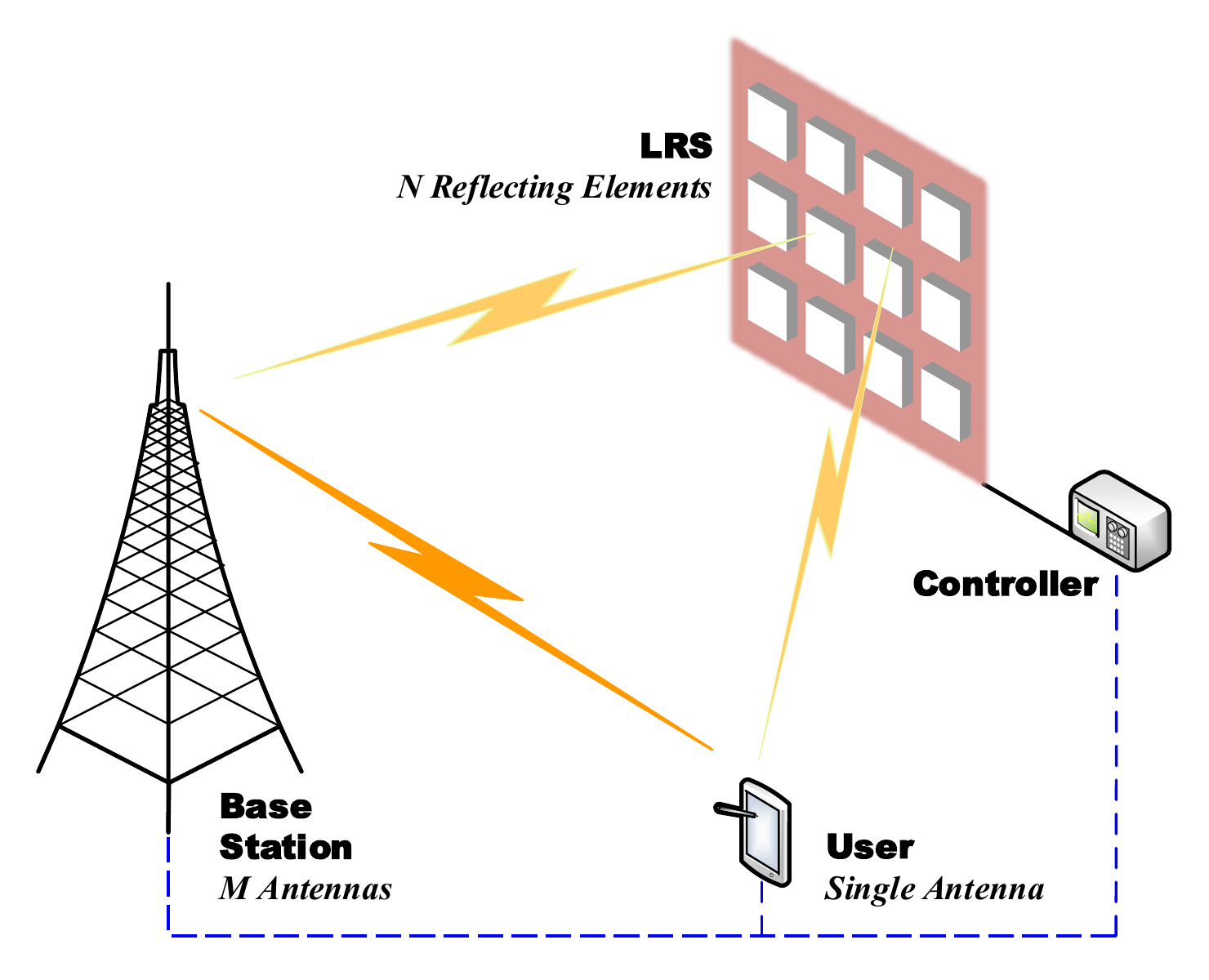}}
	\caption{The LRS-assisted wireless communication system with an $M$-antenna BS, a single-antenna user, and an LRS comprising $N$ reflecting elements.}
	\label{fig1}
\end{figure}

We consider an LRS-assisted wireless communication system in this paper, as illustrated in Fig.~\ref{fig1}. The system consists of an $M$-antenna BS, an LRS comprising $N$ reflecting elements, and a single-antenna user. In this section, we give the communication system model based on the physically correct system models in prior works \cite{8936989, 8930608, 8888223, 8647620}. The operations at the LRS is represented by the diagonal matrix $\mathbf{\Phi} = \operatorname{diag}\left(e^{j\theta_{1}},\cdots,e^{j\theta_{N}}\right)$, where $\theta_{i} \in [0,2\pi]$ represents the phase-shift of the $i^{th}$ reflecting element. The channel realizations are generated randomly and are independent between blocks, which basically covers all physical channel distributions. Denote the channels of BS-user link, BS-LRS link and LRS-user link as $\mathbf{h}_{\mathrm{d}} \in \mathbb{C}^{M\times1}$, $\mathbf{G} \in \mathbb{C}^{M \times N}$ and $\mathbf{h}_{\mathrm{r}} \in \mathbb{C}^{N \times 1}$, respectively. They are modeled as ergodic processes with fixed independent realizations, $\mathbf{h}_{\mathrm{d}} \sim \mathcal{C}\mathcal{N} \left(0, \mathbf{C}_{\mathrm{d}} \right)$ and $\mathbf{H}_{\mathrm{LRS}}=\mathbf{G}\operatorname{diag}\left(\mathbf{h}_{\mathrm{r}}\right) \sim \mathcal{C}\mathcal{N}\left(0,\mathbf{C}_{\mathrm{LRS}}\right)$, where $\mathcal{C}\mathcal{N}(\cdot)$ represents a circularly symmetric complex Gaussian distribution, and $\mathbf{C}_{\mathrm{d}}$, $\mathbf{C}_{\mathrm{LRS}}$ are the positive semi-definite covariance matrices.

\begin{figure}[htbp]
	\vspace{-1.25 em}
	\centerline{\includegraphics[width=7.75cm]{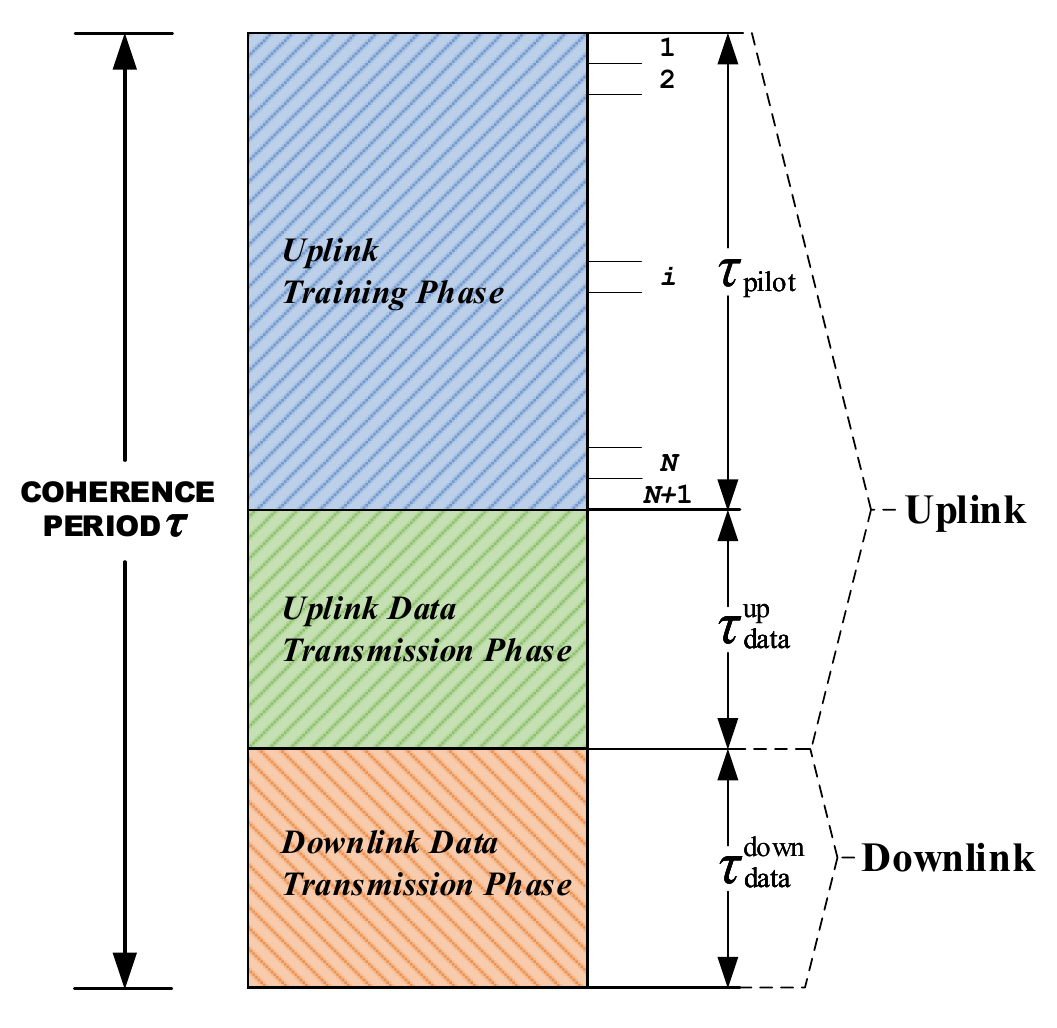}}
	\caption{The communication protocol that we adopt for the LRS-assisted wireless communication system.}
	\label{fig2}
\end{figure}

The communication protocol we adopt for the LRS-assisted system in this paper is based on the protocol proposed in \cite{nadeem2019intelligent}, as illustrated in Fig.~\ref{fig2}. The channel coherence period $\tau$ is divided into three phases: an uplink training phase of $\tau_{\mathrm{pilot}}$, an uplink transmission phase of $\tau_{\mathrm{data}}^{\mathrm{up}}$, and a downlink transmission phase of $\tau_{\mathrm{data}}^{\mathrm{down}}$. During the uplink training phase, the deterministic pilot signal $x$ is transmitted by the user to estimate channels, where the average power of $x$ is $\mathbb{E}\left\{|x|^{2}\right\} = p_{\mathrm{UE}}$. Since the LRS has no radio resources to transmit pilot signals, the BS has to estimate the cascaded channel of $\mathbf{G}$ and $\mathbf{h}_{\mathrm{r}}$, which is defined as $\mathbf{H}_{\mathrm{LRS}} =\mathbf{G}\operatorname{diag}\left(\mathbf{h}_{\mathrm{r}}\right)= \left[\mathbf{h}_{1}, \cdots, \mathbf{h}_{N}\right]$. Each column vector $\mathbf{h}_{i} \sim \mathcal{C}\mathcal{N}\left(0,\mathbf{C}_{i}\right)$ in $\mathbf{H}_{\mathrm{LRS}}$ represents the channel between the BS and the user through LRS when only the $i^{th}$ reflecting element is ON. The uplink training phase is divided into $(N+1)$ subphases. During the $1^{st}$ subphase, all reflecting elements are OFF and the BS estimates the direct channel $\mathbf{h}_{\mathrm{d}}$; During the $(i+1)^{th}$ subphase, only the $i^{th}$ reflecting element is ON and the BS estimates the channel $\mathbf{h}_{i}$. By exploiting channel reciprocity, the BS will transmit data to the user during the downlink transmission phase.

The aggregate hardware impairments of transceiver can be modeled as independent additive distortion noises \cite{5456453, zetterberg2011experimental}. The distortion noise at the user $\eta_{\mathrm{UE}} \in \mathbb{C}$ obeys the distribution of $\mathcal{C} \mathcal{N}\left(0, v_{\mathrm{UE}}\right)$, and the distortion noise at the BS $\boldsymbol{\eta}_{\mathrm{BS}} \in \mathbb{C}^{M\times1} $ obeys the distribution of $\mathcal{C N}\left(\mathbf{0}, \mathbf{\Upsilon}_{\mathrm{BS}}\right)$, where $v_{\mathrm{UE}}$ and $\mathbf{\Upsilon}_{\mathrm{BS}}$ are the variance/covariance matrix of the distortion noise. The distortion noise at an antenna is proportional to the signal power at this antenna \cite{5456453, zetterberg2011experimental}, thus we have:

\begin{itemize}	
	\item During the $1^{st}$ subphase of uplink training phase, the distortion noise covariance matrix $\mathbf{\Upsilon}_{\mathrm{BS}}$ can be modeled as $\mathbf{\Upsilon}_{\mathrm{BS}}=\kappa_{\mathrm{BS}}\left(p_{\mathrm{UE}}+\kappa_{\mathrm{UE}}p_{\mathrm{UE}}\right)\operatorname{diag}\left(\mathbf{C}_{\mathrm{d}}\right)$, where $\kappa_{\mathrm{UE}}$ and $\kappa_{\mathrm{BS}}$ are respectively the proportionality coefficients which characterize the levels of hardware impairments at the user and the BS, and are related to the error vector magnitude (EVM). The EVM is a common measure of hardware quality for transceivers, \textit{e.g.}, when the BS transmits the signal $\mathbf{x}$, the EVM at the BS is defined as
	\begin{equation}
	\mathrm{EVM}_{\mathrm{BS}} = \sqrt{\frac{\operatorname{tr}\left( \mathbb{E}\left\{\boldsymbol{\eta}_{\mathrm{BS}} \boldsymbol{\eta}_{\mathrm{BS}}^{\mathrm{H}}\right\} \right) }{\operatorname{tr}\left( \mathbb{E}\left\{\mathbf{x} \mathbf{x}^{\mathrm{H}}\right\}\right) }} = \sqrt{\kappa_{\mathrm{BS}}}.
	\end{equation}
		
	\item During the $(i+1)^{th}$ subphase of uplink training phase, the distortion noise covariance matrix $\mathbf{\Upsilon}_{\mathrm{BS}}$ can be modeled as $\mathbf{\Upsilon}_{\mathrm{BS}}=\kappa_{\mathrm{BS}}\left(p_{\mathrm{UE}}+\kappa_{\mathrm{UE}}p_{\mathrm{UE}}\right)\operatorname{diag}\left(\mathbf{C}_{\mathrm{d}}+\mathbf{C}_{i}\right)$.	
	
	\item During the uplink data transmission phase, the distortion noise covariance matrix $\mathbf{\Upsilon}_{\mathrm{BS}}$ can be modeled as $\mathbf{\Upsilon}_{\mathrm{BS}}=\kappa_{\mathrm{BS}}\left(p_{\mathrm{UE}}+\kappa_{\mathrm{UE}}p_{\mathrm{UE}}\right)\operatorname{diag}\left(\mathbf{C}_{\mathrm{d}}+\sum_{i=1}^{N}\mathbf{C}_{i}\right)$.	
	
	\item During the uplink training phase as well as the uplink data transmission phase, the distortion noise variance $v_{\mathrm{UE}}$ can be modeled as $v_{\mathrm{UE}}=\kappa_{\mathrm{UE}} p_{\mathrm{UE}}$.	
\end{itemize}

The hardware impairments of LRS can be modeled as phase noise since the LRS is a passive device and high-precision configuration of the reflection phases is infeasible. The phase noise of the $i^{th}$ element of LRS is denoted as $\Delta \theta_{i}$, which is randomly distributed on $[-\pi, \pi)$ according to a certain circular distribution. Due to the reasonable assumption in \cite{8869792}, the distribution of the phase noise $\Delta \theta_{i}$ has mean direction zero, \textit{i.e.}, $\arg \left(\mathbb{E}\left\{e^{j \Delta \theta_{i}}\right\}\right)=0$, and its probability density function is symmetric around zero. The actual matrix of LRS with phase noise is $\widetilde{\mathbf{\Phi}}=\operatorname{diag}\left(e^{j(\theta_{1}+\Delta\theta_{1})}, e^{j(\theta_{2}+\Delta\theta_{2})}, \cdots, e^{j(\theta_{N}+\Delta\theta_{N}}\right)$.

Based on the communication system model given above, the received pilot signals $\mathbf{y}_{\mathrm{d}},\mathbf{y}_{1},\cdots,\mathbf{y}_{N} \in\mathbb{C}^{M\times1}$ at the BS in different subphases of uplink training phase are
\begin{equation}
\label{eq.1}
\mathbf{y}_{\mathrm{d}}=\mathbf{h}_{\mathrm{d}}\left(x+\eta_{\mathrm{UE}}\right)+\boldsymbol{\eta}_{\mathrm{BS}}+\mathbf{n},
\end{equation}
\begin{equation}
\label{eq.2}
\mathbf{y}_{i}=\mathbf{h}_{\mathrm{d}}\left(x+\eta_{\mathrm{UE}}\right)+\mathbf{h}_{i}\left[e^{j\Delta\theta_{i}}\left(x+\eta_{\mathrm{UE}}\right)\right]+\boldsymbol{\eta}_{\mathrm{BS}}+\mathbf{n},
\end{equation}
where $x \in \mathbb{C}$ is the deterministic pilot signal, and $\mathbf{n} \in \mathbb{C}^{M \times 1}$ is an additive white Gaussian noise with the elements independently drawn from $\mathcal{C} \mathcal{N}\left(0, \sigma_{\mathrm{BS}}^{2}\right)$. The received signal $\mathbf{y} \in\mathbb{C}^{M\times1}$ at the BS during the uplink data transmission phase from the user is
\begin{equation}
\mathbf{y}=\left(\mathbf{h}_{\mathrm{d}}+\mathbf{G} \widetilde{\mathbf{\Phi}} \mathbf{h}_{\mathrm{r}}\right)\left(x+\eta_{\mathrm{UE}}\right)+\boldsymbol{\eta}_{\mathrm{BS}}+\mathbf{n},
\end{equation}
where $x \in \mathbb{C}$ is the transmitted data signal, and the transmit power is $ p_{\mathrm{UE}} = \mathbb{E}\left\{|x|^{2}\right\} $ which is same with the pilot power.

\section{Channel Estimation Performance}
In this section, we analyze the channel estimation performance of the LRS system with LMMSE estimator. The estimated channels include direct channel $\mathbf{h}_{\mathrm{d}}$ and column vectors $\mathbf{h}_{i}$ of cascaded channel $\mathbf{H}_{\mathrm{LRS}}$. When we estimate the direct channel in the $1^{st}$ subphase, all reflecting elements of LRS are OFF. The system can be simplified as a multiple-input single-output (MISO) communication system. The corresponding estimation performance was given in Theorem 1 of \cite{6891254}, which is shown in Lemma 1 as follows. 

\begin{lemma}
	The estimated direct channel $ \hat{\mathbf{h}}_{\mathrm{d}} $ using LMMSE estimator can be represented as
	\begin{equation}
	\hat{\mathbf{h}}_{\mathrm{d}} = x^{*} \mathbf{C}_{\mathrm{d}} \mathbf{Y}_{\mathrm{d}}^{-1} \mathbf{y}_{\mathrm{d}},
	\end{equation}
	where $\mathbf{Y}_{\mathrm{d}}$ is the covariance matrix of the received pilot signal $\mathbf{y}_{\mathrm{d}}$. The LMMSE is the trace of the error covariance matrix, $\operatorname{tr}\left( \mathbf{M}_{\mathrm{d}}\right) $, and $\mathbf{M}_{\mathrm{d}}$ is
	\begin{equation}
	\mathbf{M}_{\mathrm{d}} = \mathbf{C}_{\mathrm{d}}-p_{\mathrm{UE}} \mathbf{C}_{\mathrm{d}} \mathbf{Y}_{\mathrm{d}}^{-1} \mathbf{C}_{\mathrm{d}}.
	\end{equation}
\end{lemma}

When we estimate the LRS channel $\mathbf{h}_{i}$, one important difference from the direct channel $\mathbf{h}_\mathrm{d}$ is that there exist hardware impairments on LRS, and these impairments should be taken into consideration. Another important difference is that the signal received at the BS in the $(i+1)^{th}$ subphase consists of two parts: the signal transmitted through direct channel and the signal transmitted through LRS channel. The signal ${\widetilde{\mathbf{y}}}_{i}$ transmitted through LRS channel can be obtained by subtracting the signal $\mathbf{y}_{\mathrm{d}}$ in Eq. (\ref{eq.1}) from the signal $ \mathbf{y}_{i} $ in Eq. (\ref{eq.2}), as given by
\begin{equation}
\label{eq.6}
{{\widetilde{\mathbf{y}}}_{i}} = \mathbf{h}_{i}\left[e^{j \Delta \theta_{i}}\left(x+\eta_{\mathrm{UE}}\right)\right]+\boldsymbol{\eta}_{\mathrm{BS}}^{\mathrm{LRS}}+\mathbf{n}+\mathbf{n}. 
\end{equation}
It should be noted that additive Gaussian noise cannot be eliminated, and the noise term in ${\widetilde{\mathbf{y}}}_{i}$ is the superposition of that in $ \mathbf{y}_{\mathrm{d}} $ and $ \mathbf{y}_{i} $, which still obeys a Gaussian distribution. Similarly, the power of residual distortion noise caused by hardware impairments is superposed: the distortion noise at the BS in Eq. (\ref{eq.6}) is $\boldsymbol{\eta}_{\mathrm{BS}}^{\mathrm{LRS}} \sim \mathcal{CN}\left(\mathbf{0}, \mathbf{\Upsilon}_{\mathrm{BS}}^{\mathrm{LRS}}\right)$ where $\mathbf{\Upsilon}_{\mathrm{BS}}^{\mathrm{LRS}}=\kappa_{\mathrm{BS}}\left(p_{\mathrm{UE}}+\kappa_{\mathrm{UE}}p_{\mathrm{UE}}\right)\operatorname{diag}\left(2\mathbf{C}_{\mathrm{d}}+\mathbf{C}_{i}\right)$. In addition, we omit the superposition of $\mathbf{h}_{\mathrm{d}}\eta_{\mathrm{UE}}$ in Eqs. (\ref{eq.1}) and (\ref{eq.2}) since the value of it is very small in practice.

\begin{theorem}
	The estimated LRS channel $\hat{\mathbf{h}}_{i}$ from the separated signal ${\widetilde{\mathbf{y}}}_{i}$ using LMMSE estimator is
	\begin{equation}
	\label{eq.7}
	\hat{\mathbf{h}}_{i} = x^{*}\mathbf{C}_{i}{\widetilde{\mathbf{Y}}_{i}}^{-1} {\widetilde{\mathbf{y}}}_{i},
	\end{equation}
	where $ \widetilde{\mathbf{Y}}_{i} $ is the covariance matrix of the separated signal $ \widetilde{\mathbf{y}}_{i} $. The LMMSE is the trace of the error covariance matrix, $\operatorname{tr}\left(\mathbf{M}_{i}\right) $, and $\mathbf{M}_{i}$ is
	\begin{equation}
	\label{eq.8}
	\mathbf{M}_{i} = \mathbf{C}_{i}-p_{\mathrm{UE}} \mathbf{C}_{i} \widetilde{\mathbf{Y}}_{i}^{-1} \mathbf{C}_{i}.
	\end{equation}
\end{theorem}
\begin{IEEEproof}  
	The estimated LRS channel $ \hat{\mathbf{h}}_{i} $ using LMMSE estimator has a form of $ \hat{\mathbf{h}}_{i} = \mathbf{A} \widetilde{\mathbf{y}}_{i} $, where $\mathbf{A}$ is the detector matrix which minimizes the mean square error (MSE). According to the definition of MSE, we obtain that MSE is the trace of the error covariance matrix, $\operatorname{tr}\left(\mathbf{M}_{i}\right) $, and $\mathbf{M}_{i}$ is
	\begin{equation}
	\label{eq.9}
	\mathbf{M}_{i} = \mathbb{E}\left\{\mathbf{A} \widetilde{\mathbf{y}}_{i} {{\widetilde{\mathbf{y}}}_{i}}^{\mathrm{H}} \mathbf{A}^{\mathrm{H}}+\mathbf{h}_{i} \mathbf{h}_{i}^{\mathrm{H}}-\mathbf{h}_{i} {{\widetilde{\mathbf{y}}}_{i}}^{\mathrm{H}} \mathbf{A}^{\mathrm{H}}-\mathbf{A} \mathbf{y}_{i} {{\widetilde{\mathbf{y}}}_{i}}^{\mathrm{H}}\right\}.
	\end{equation}
	By substituting $ \widetilde{\mathbf{y}}_{i} $ in Eq. (\ref{eq.6}) into Eq. (\ref{eq.9}), we obtain that
	\begin{equation}
	\label{eq.10}
	\text{MSE}=\operatorname{tr}\left(\mathbf{A} \widetilde{\mathbf{Y}}_{i} \mathbf{A}^{\mathrm{H}} - x \mathbf{A} \mathbf{C}_{i} - x^{*} \mathbf{C}_{i} \mathbf{A}^{\mathrm{H}} + \mathbf{C}_{i} \right).
	\end{equation}
	Then, the detector matrix $\mathbf{A}$ which minimizes the MSE can be obtained by equaling the derivative of Eq. (\ref{eq.10}) with respect to $ \mathbf{A} $ to zero, as given by
	\begin{equation}
	\frac{\partial \text{MSE}}{\partial \mathbf{A}}=\mathbf{0} \Rightarrow \mathbf{A} = x^{*} \mathbf{C}_{i} \widetilde{\mathbf{Y}}_{i}^{-1} .
	\end{equation}
	Finally, we obtain the estimated LRS channel $\hat{\mathbf{h}}_{i}$ in Eq. (\ref{eq.7}). By substituting $ \mathbf{A} $ into Eq. (\ref{eq.9}), we obtain the error covariance matrix $\mathbf{M}_{i}$ in Eq. (\ref{eq.8}). 
\end{IEEEproof}

\begin{remark}
	The phase errors of the reflecting elements are random and unknown to the BS in practice. We can only use the statistic characteristics of $\Delta \theta_{i}$ to estimate the LRS channel. The result shows that the LRS hardware impairments will not affect the estimation accuracy statistically. Thus, a massive MIMO system can be replaced by an LRS-assisted system with large number of low-quality reflecting elements and moderate number of high-quality antennas, which causes tolerable decrease of estimation accuracy but can reduce hardware cost substantially. In addition, the estimation accuracy will decrease on account of the superposition of noise/distortion power caused by the subtraction operation on signals, and we need more accurate estimation method to compensate this loss.  
\end{remark}

\begin{corollary}
	The average estimation error per antenna is independent of the number of BS antennas, but correlated to the number of reflecting elements on LRS (the times of estimation increases with the number of reflecting elements). Contrary to the ideal hardware case that the error variance converges to zero as $ p_{\mathrm{UE}}\rightarrow \infty $, the transceiver hardware impairments limit the estimation performance.
\end{corollary}

\begin{IEEEproof}  
	Consider the special case of $ \mathbf{C}_{\mathrm{d}}=\lambda \mathbf{I} $ and $ \mathbf{C}_{i}=\lambda \mathbf{I} $. 
	The covariance matrix of the direct channel estimation error is 
	\begin{equation}
	\mathbf{M}_{\mathrm{d}}=\lambda\mathbf{I}-\frac{\lambda^2}{\lambda\kappa_{\mathrm{d}}+\frac{\sigma_{\mathrm{BS}}^{2}}{p_{\mathrm{UE}}}}\mathbf{I},
	\end{equation}
	where $\kappa_{\mathrm{d}} = 1+\kappa_{\mathrm{UE}}+\kappa_{\mathrm{BS}}\left(1+\kappa_{\mathrm{UE}}\right)$.
	The covariance matrix of the estimation error of the channel through the $i^{th}$ element of LRS is
	\begin{equation}
	\mathbf{M}_{i}=\lambda\mathbf{I}-\frac{\lambda^2}{\lambda\kappa_{i}+2\frac{\sigma_{\mathrm{BS}}^{2}}{p_{\mathrm{UE}}}}\mathbf{I},
	\end{equation}
	where $ \kappa_{i} = 1+\kappa_{\mathrm{UE}}+3\kappa_{\mathrm{BS}}\left(1+\kappa_{\mathrm{UE}}\right) $. In the high pilot signal power regime, we have
	\begin{equation}
	\lim _{p_{\mathrm{UE}} \rightarrow \infty} \mathbf{M}_{\mathrm{d}}=\lambda\mathbf{I}-\frac{\lambda}{\kappa_{\mathrm{d}}}\mathbf{I}, \text{ and } \lim _{p_{\mathrm{UE}} \rightarrow \infty} \mathbf{M}_{i}=\lambda\mathbf{I}-\frac{\lambda}{\kappa_{i}}\mathbf{I}.
	\end{equation}
	Thus, perfect estimation accuracy cannot be achieved in practice, not even asymptotically.
\end{IEEEproof}

\begin{figure}[htbp]
	\vspace{-1 em}
	\centerline{\includegraphics[width=8.7cm]{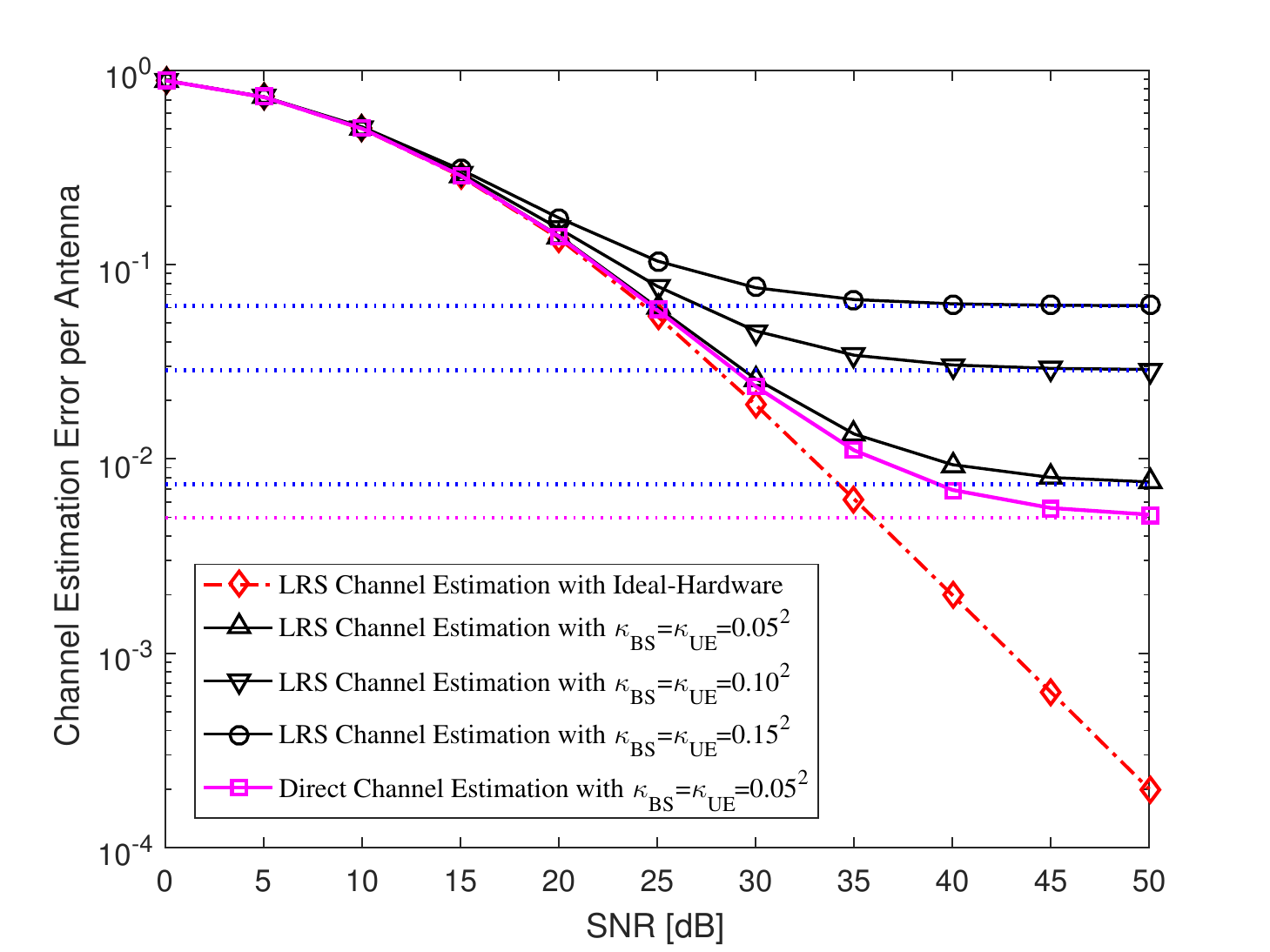}}
	\caption{Channel estimation error per antenna of direct channel and LRS channel with ideal and non-ideal hardware.}
	\label{fig3}
\end{figure}

We compare the estimation performance of direct channel and LRS channel with different impairment levels to illustrate the difference between them as well as the estimation accuracy limit caused by hardware impairments. We assume that the number of BS antennas is $M=20$, and the hardware impairments coefficients are chosen from the set of $\left\lbrace 0, 0.05^2, 0.10^2, 0.15^2\right\rbrace$. The channel covariance matrix is generated by the exponential correlation model from \cite{951380}. Fig.~\ref{fig3} shows the channel estimation error per antenna averaged by the trace of $\mathbf{C}_{i}$ (in the case of direct channel, it is averaged by the trace of $\mathbf{C}_{\mathrm{d}}$), and it is a decreasing function of the average SNR which is defined as $ \text{SNR} = \frac{p_{\mathrm{UE}} \operatorname{tr}\left( \mathbf{C}_{i}\right)}{M\sigma_{\mathrm{BS}}^{2}}$. We notice that the estimation error increases with the impairment level, and hardware impairments create non-zero error floors. In addition, the estimation error of LRS channel is larger than that of direct channel.

\begin{figure}[htbp]
	\vspace{-1 em}
	\centerline{\includegraphics[width=8.7cm]{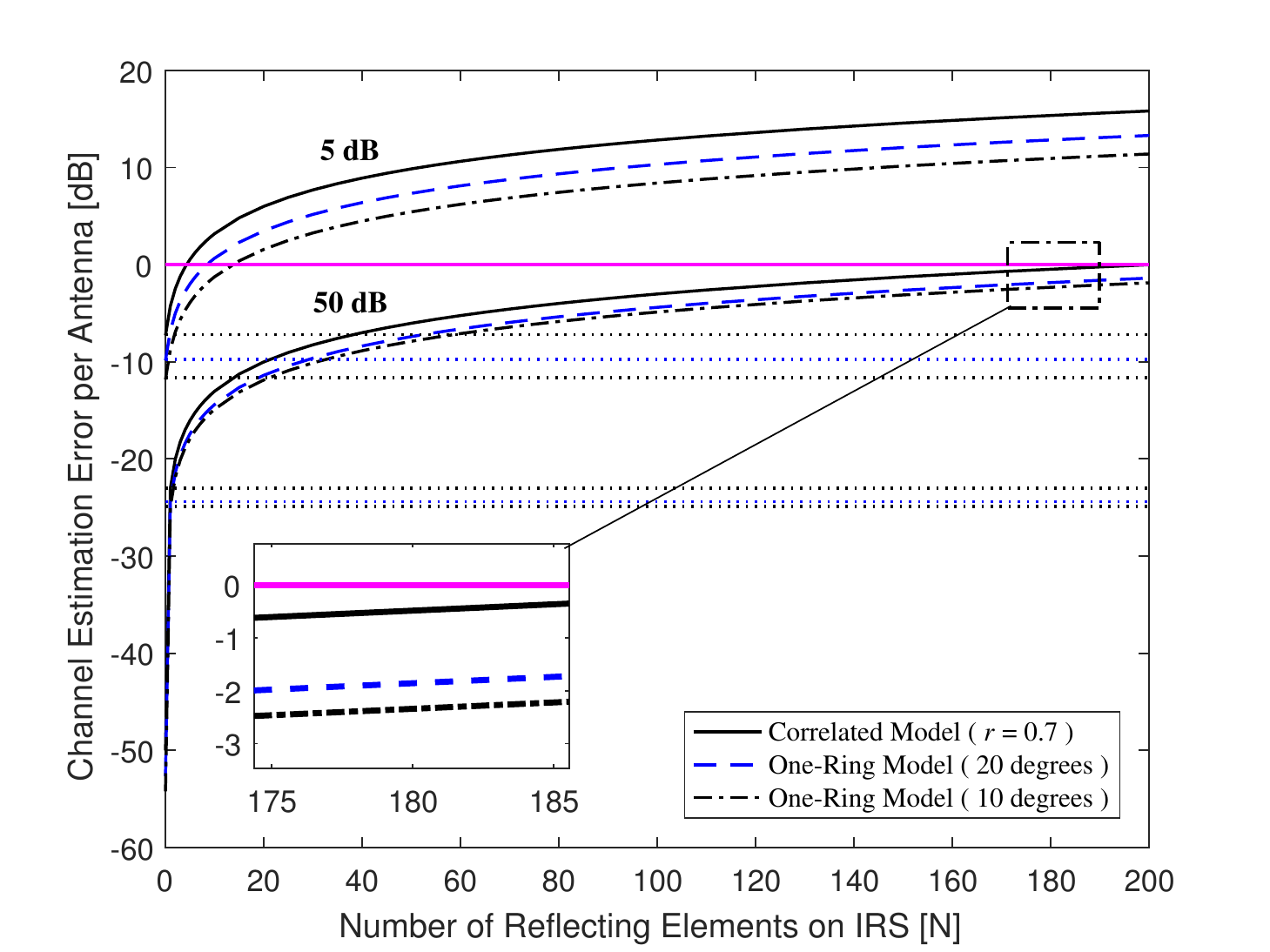}}
	\caption{Channel estimation error per antenna versus the number of reflecting elements for $ \text{SNR}=5 \text{ dB} \text{ and } 50 \text{ dB} $. Three channel covariance models are considered and $ \kappa_{\mathrm{BS}} = \kappa_{\mathrm{UE}} = 0.05^2 $.}
	\label{fig4}
\end{figure}

To numerically illustrate the effect of different numbers of reflecting elements on channel estimation performance, we assume the number ranges from 0 to 200. We consider three models to generate the channel covariance matrix: 1) Exponential correlation model with correlation coefficient $r=0.7$ \cite{951380}; 2) One-ring model with 20 degrees angular spread; 3) One-ring model with 10 degrees angular spread \cite{837052}. Fig.~\ref{fig4} shows that the channel estimation error increases with the increase of the number of reflecting elements and decreases with the increase of SNR. We notice that the estimation error is less than 0 dB with large number of reflecting elements when SNR is over 50 dB.

\section{Power Scaling Law of User}
Many related works \cite{6415388, 6457363, 6902790} show that the emitted power can be reduced with no reduction in performance by utilizing the array gain in multi-antenna  system. One can reduce the transmit power as $1 / M^{\alpha}$, $0<\alpha<\frac{1}{2}$, and still achieve non-zero spectral efficiency as $M \rightarrow \infty$. In this section, we quantify the power scaling law for LRS-assisted wireless communication system. By considering maximum-ratio combining (MRC) detector as it achieves fairly well performance \cite{5595728, 6457363}, we consider the cases of perfect channel state information and estimated channel state information with error. The received signal at the BS with non-ideal hardware is $\mathbf{y}=(\mathbf{h}_{\mathrm{d}}+\mathbf{G} \widetilde{\mathbf{\Phi}} \mathbf{h}_{\mathrm{r}})(x+\eta_{\mathrm{UE}})+\boldsymbol{\eta}_{\mathrm{BS}}+\mathbf{n}$, where $\mathbf{h}_{\mathrm{d}}$, $\mathbf{G}$ and $\mathbf{h}_{\mathrm{r}}$ are mutually independent matrices whose elements are i.i.d. zero-mean random variables. According to the law of large numbers, we have 
\begin{equation}
\label{eq.15}
\frac{1}{M} \mathbf{h}_{\mathrm{d}}^{\mathrm{H}} \mathbf{h}_{\mathrm{d}} \rightarrow \sigma_{\mathrm{d}}^{2}, \text { as } M \rightarrow \infty,
\end{equation}
where $ \sigma_{\mathrm{d}}^{2} = \mathbb{E}\left\lbrace \left|{h}_{\mathrm{d},i}\right|^{2}\right\rbrace $ and $ {h}_{\mathrm{d},i} $ is the element of the channel vector $ \mathbf{h}_{\mathrm{d}} $.
According to the rule of matrix multiplication, we obtain
\begin{equation}
\frac{1}{N} \mathbf{G} \mathbf{h}_{\mathrm{r}} \rightarrow \left[\mathbb{E}\left\lbrace {G}_{i,j} {h}_{\mathrm{r},j}\right\rbrace ,\cdots,\mathbb{E}\left\lbrace {G}_{i, j} {h}_{\mathrm{r}, j}\right\rbrace \right]^{\mathrm{T}},\text{ as } N \rightarrow \infty,
\end{equation}
where $ {G}_{i,j} $ is the element of channel matrix $  \mathbf{G} $, and $ {h}_{\mathrm{r},j} $ is the element of channel vector $ \mathbf{h}_{\mathrm{r}} $. As $ \mathbf{G} \mathbf{h}_{\mathrm{r}}/N $ is a random vector similar to $ \mathbf{h}_{\mathrm{d}} $, we reuse Eq. (\ref{eq.15}) to obtain that
\begin{equation}
\label{eq.17}
\frac{1}{MN^{2}}\left(\mathbf{G} \mathbf{h}_{\mathrm{r}}\right)^{\mathrm{H}}\left(\mathbf{G} \mathbf{h}_{\mathrm{r}}\right) \rightarrow \sigma_{\mathrm{LRS}}^{2}, \text{ as } M,N \rightarrow \infty,
\end{equation}
where $ \sigma_{\mathrm{LRS}}^{2} = \mathbb{E}\left\lbrace \left|{G}_{i, j} {h}_{\mathrm{r}, j}\right|^{2}\right\rbrace $.

\subsubsection{BS with perfect channel state information}
We first consider the case where the BS can obtain perfect channel state information. The detector vector is $\mathbf{A} = \mathbf{h}_{\mathrm{d}}+\mathbf{G}\mathbf{\Phi}\mathbf{h}_{\mathrm{r}}$ when using MRC. As illustrated in Section III, the phase error of LRS is random and unknown to the BS, thus the MRC detector is $ \mathbf{h}_{\mathrm{d}}+\mathbf{G}\mathbf{\Phi}\mathbf{h}_{\mathrm{r}} $ rather than $ \mathbf{h}_{\mathrm{d}}+\mathbf{G}\widetilde{\mathbf{\Phi}}\mathbf{h}_{\mathrm{r}} $. The transmitted signal can be detected by multiplying the received signal $\mathbf{y} $ with $ \mathbf{A}^{\mathrm{H}} $, \textit{i.e.}, $r=\mathbf{A}^{\mathrm{H}}\mathbf{y}$. The received signal after using the detector vector $ \mathbf{\mathbf{A}} $ is given as
\begin{equation}
r=\mathbf{h}^{\mathrm{H}}\widetilde{\mathbf{h}}\left(x+\eta_{\mathrm{UE}}\right)+\mathbf{h}^{\mathrm{H}}\boldsymbol{\eta}_{\mathrm{BS}}+\mathbf{h}^{\mathrm{H}}\mathbf{n},
\end{equation}
where, for simplicity, $\mathbf{h}$ represents $\mathbf{h}_{\mathrm{d}}+\mathbf{G}\mathbf{\Phi}\mathbf{h}_{\mathrm{r}}$ and $\widetilde{\mathbf{h}}$ represents $\mathbf{h}_{\mathrm{d}}+\mathbf{G}\widetilde{\mathbf{\Phi}}\mathbf{h}_{\mathrm{r}}$. In addition, the phase noise on LRS will not change the signal power, and the expectation of $ \Delta \theta_{i} $ is zero. We obtain the achievable rate of uplink in Eq. (\ref{eq.19}).

\begin{figure*}[!t]
	\normalsize
	\setcounter{MYtempeqncnt}{\value{equation}}
	\begin{equation}
	\label{eq.19}
	\mathcal{R}_{\mathrm{up}}=\mathbb{E}\left\{\log _{2}\left(1+\frac{p_{\mathrm{UE}}\left|\mathbf{h}^{\mathrm{H}}\mathbf{h}\right|^{2}}{\kappa_{\mathrm{UE}}p_{\mathrm{UE}}\left|\mathbf{h}^{\mathrm{H}}\mathbf{h}\right|^{2}+\left\|\mathbf{h}\right\|_{2}^{2}\left(\sigma_{\mathrm{BS}}^{2}+p_{\mathrm{UE}}\kappa_{\mathrm{BS}}\left(1+\kappa_{\mathrm{UE}}\right)\right)}\right)\right\}
	\end{equation}
	\begin{equation}
	\label{eq.20}
	\mathcal{R}_{\mathrm{up}}=\mathbb{E}\left\{\log _{2}\left(1+\frac{p_{\mathrm{UE}}\left|\mathbf{h}_{\mathrm{est}}^{\mathrm{H}}\mathbf{h}_{\mathrm{est}}\right|^{2}}{\left(1+\kappa_{\mathrm{UE}}\right) p_{\mathrm{UE}}\left\|\mathbf{h}_{\mathrm{est}}\right\|_{2}^{2}\frac{\beta}{p_{\mathrm{UE}} \beta+1}+\kappa_{\mathrm{UE}}p_{\mathrm{UE}}\left|\mathbf{h}_{\mathrm{est}}^{\mathrm{H}}\mathbf{h}_{\mathrm{est}}\right|^{2}+\left\|\mathbf{h}_{\mathrm{est}}\right\|_{2}^{2}\left(\sigma_{\mathrm{BS}}^{2}+p_{\mathrm{UE}}\kappa_{\mathrm{BS}}\left(1+\kappa_{\mathrm{UE}}\right)\right)}\right)\right\}
	\end{equation}
	\setcounter{figure}{\value{MYtempeqncnt}}
	\hrulefill
	\vspace*{3pt}
\end{figure*} 

\begin{proposition}
	Assume that the BS has perfect channel state information and the transmit power of the user is scaled with $ M $ and $ N $ according to $p_{\mathrm{UE}}={\mathrm{E}_{\mathrm{UE}}}/{(M+k M N^{2})}$, where $ \mathrm{E}_{\mathrm{UE}} $ is fixed and $ k= {\sigma_{\mathrm{LRS}}^{2}}/{ \sigma_{\mathrm{d}}^{2}}$, we have 
	\begin{equation}
	\label{eq.21}
	\mathcal{R}_{\mathrm{up}} \rightarrow \log _{2}\left(1+\frac{\mathrm{E}_{\mathrm{UE}}\sigma_{\mathrm{d}}^{2}}{\kappa_{\mathrm{UE}}\mathrm{E}_{\mathrm{UE}}\sigma_{\mathrm{d}}^{2}+\sigma_{\mathrm{BS}}^{2}}\right), \text{ as } M, N \rightarrow \infty.
	\end{equation}
\end{proposition}

\begin{IEEEproof}
	Substituting $p_{\mathrm{UE}}={\mathrm{E}_{\mathrm{UE}}}/{\left( M+k M N^{2}\right)}$ into Eq. (\ref{eq.19}), and using the law of large numbers reviewed in Eqs. (\ref{eq.15}) and (\ref{eq.17}), we obtain the convergence value of the achievable rate as $M, N \rightarrow \infty$ in Eq. (\ref{eq.21}).
\end{IEEEproof}

\subsubsection{BS with imperfect channel state information}
In practice, the BS has to estimate the channel, and there exists estimation error as we discussed in Section III. For simplicity, we denote estimation error as $\boldsymbol{\mathcal{E}}=\mathbf{h}_{\mathrm{est}}-\widetilde{\mathbf{h}}$. Referring to the Eq. (33) in \cite{6457363}, the elements of $\boldsymbol{\mathcal{E}}$ are random variables with zero means and variances $\frac{\beta}{p_{\mathrm{UE}} \beta+1}$, where $\beta=\left(1+k N^{2}\right)\sigma_{\mathrm{d}}^{2}$. The received signal can be rewritten as 
\begin{equation}
r= \mathbf{h}_{\mathrm{est}}^{\mathrm{H}}\left( \mathbf{h}_{\mathrm{est}}-\boldsymbol{\mathcal{E}}\right) \left(x+\eta_{\mathrm{UE}}\right)+\mathbf{h}_{\mathrm{est}}^{\mathrm{H}}\boldsymbol{\eta}_{\mathrm{BS}}+\mathbf{h}_{\mathrm{est}}^{\mathrm{H}}\mathbf{n}.
\end{equation}
Similar to the Eq. (38) in \cite{6457363}, the achievable rate of uplink channel is given in Eq. (\ref{eq.20}), where each element of $\mathbf{h}_{\mathrm{est}}^{\mathrm{H}}$ is a random variable with zero mean and variance $\frac{p_{\mathrm{UE}} \beta^{2}}{p_{\mathrm{UE}} \beta+1}$.

\begin{proposition}
	Assume that the BS has imperfect channel state information and the transmit power of the user is scaled with $ M $ and $ N $ according to $p_{\mathrm{UE}}={\mathrm{E}_{\mathrm{UE}}}/{(\sqrt{M}(1+k N^{2})) }$, where $ \mathrm{E}_{\mathrm{UE}} $ is fixed and $ k= {\sigma_{\mathrm{LRS}}^{2}}/{ \sigma_{\mathrm{d}}^{2}}$, we have 
	\begin{equation}
	\label{eq.23}
	\mathcal{R}_{\mathrm{up}} \rightarrow \log _{2}\left(1+\frac{\mathrm{E}^{2}_{\mathrm{UE}}\sigma_{\mathrm{d}}^{4}}{\kappa_{\mathrm{UE}}\mathrm{E}^{2}_{\mathrm{UE}}\sigma_{\mathrm{d}}^{4}+\sigma_{\mathrm{BS}}^{2}}  \right), \text{ as } M, N \rightarrow \infty.
	\end{equation}
\end{proposition}

\begin{IEEEproof}
	The proof follows the similar procedures with Proposition 1. Substituting $p_{\mathrm{UE}}={\mathrm{E}_{\mathrm{UE}}}/{(\sqrt{M}(1+k N^{2}))}$ into Eq. (\ref{eq.20}), and using the law of large number reviewed in Eqs. (\ref{eq.15}) and (\ref{eq.17}) along with the variances of elements of estimation error vector $ \boldsymbol{\mathcal{E}} $ and channel estimation vector $\mathbf{h}_{\mathrm{est}} $, we obtain the convergence value of the achievable rate as $M, N \rightarrow \infty$ in Eq. (\ref{eq.23}). 
\end{IEEEproof}

\begin{remark}
	Proposition 1 shows that with perfect channel state information and a large $ M $ and $ N $, the performance of an LRS-assisted system with $ M $-antenna BS, $ N $-reflecting element LRS and the transmit power $ {\mathrm{E}_{\mathrm{UE}}}/ {(M(1+k N^{2}))} $ of the user is equal to the performance of a single-input single-output (SISO) system with transmit power $\mathrm{E}_{\mathrm{UE}}$. Proposition 2 shows that with imperfect channel state information and a large $ M $ and $ N $, the performance of an LRS-assisted system with $ M $-antenna BS, $ N $-reflecting element LRS and the transmit power $ {\mathrm{E}_{\mathrm{UE}}}/ {(\sqrt{M}(1+k N^{2}))} $ of the user is equal to the performance of a SISO system with transmit power $\mathrm{E}^{2}_{\mathrm{UE}}\sigma_{\mathrm{d}}^{2}$. Proposition 2 also implies that the transmit power can be cut proportionally to ${\mathrm{E}_{\mathrm{UE}}}/{(M^{\alpha}(1+k N^{2})^{2\alpha})}$, where $\alpha \leq \frac{1}{2}$. If $\alpha>\frac{1}{2}$, the achievable rate of uplink channel converges towards zero as $M \rightarrow \infty$ and $N \rightarrow \infty$.
\end{remark}

\begin{figure}[htbp]
	\vspace{-1 em}
	\setcounter{figure}{4}
	\centerline{\includegraphics[width=8.7cm]{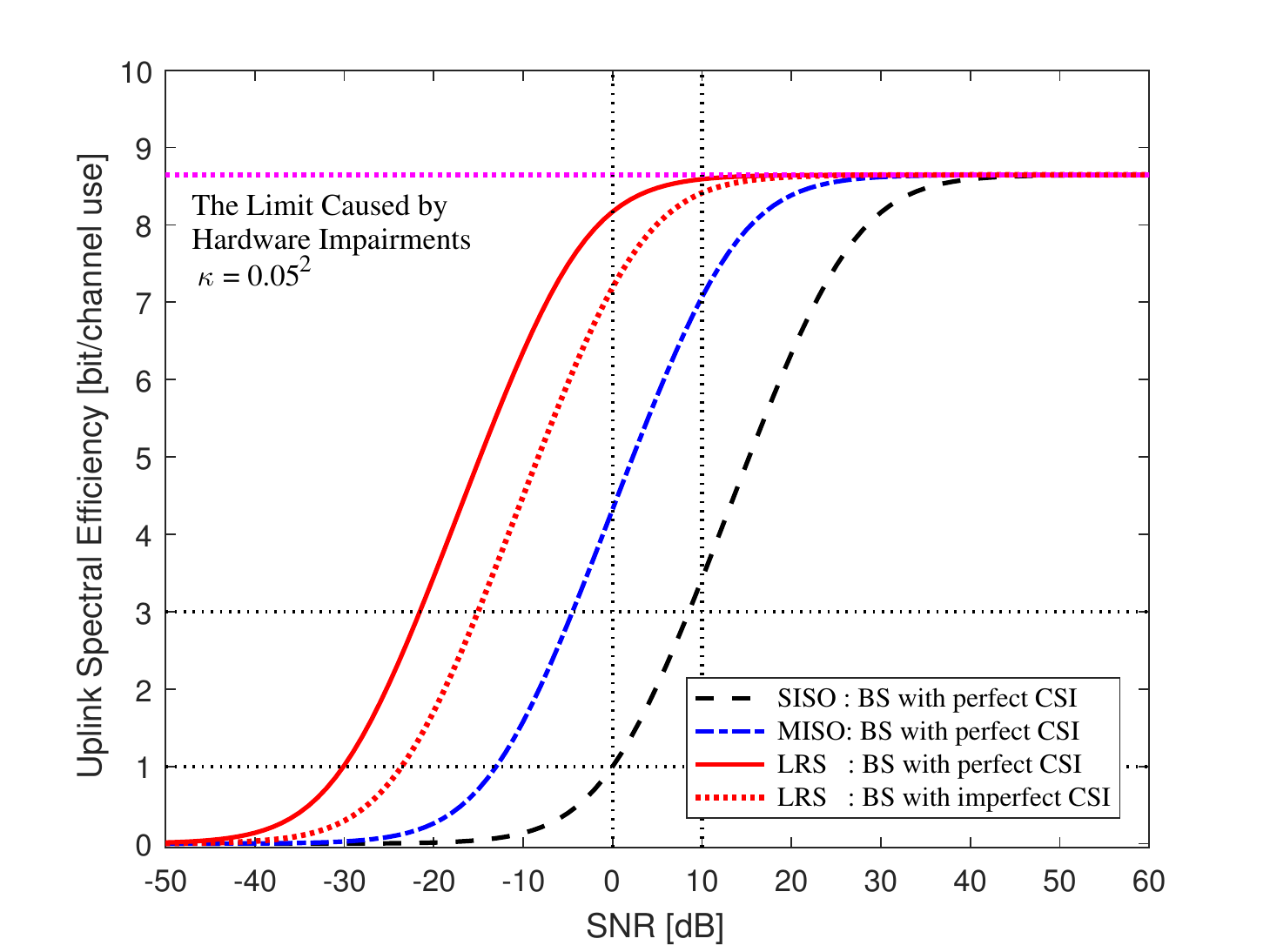}}
	\caption{Spectral efficiency on the uplink versus SNR for $ M=20 $, $ N=100 $ and $  \kappa_{\mathrm{BS}}=\kappa_{\mathrm{UE}}=0.05^2 $ with perfect and imperfect channel state information.}
	\label{fig5}
\end{figure}

\begin{figure}[htbp]
	\vspace{-0.6 em}
	\centerline{\includegraphics[width=8.7cm]{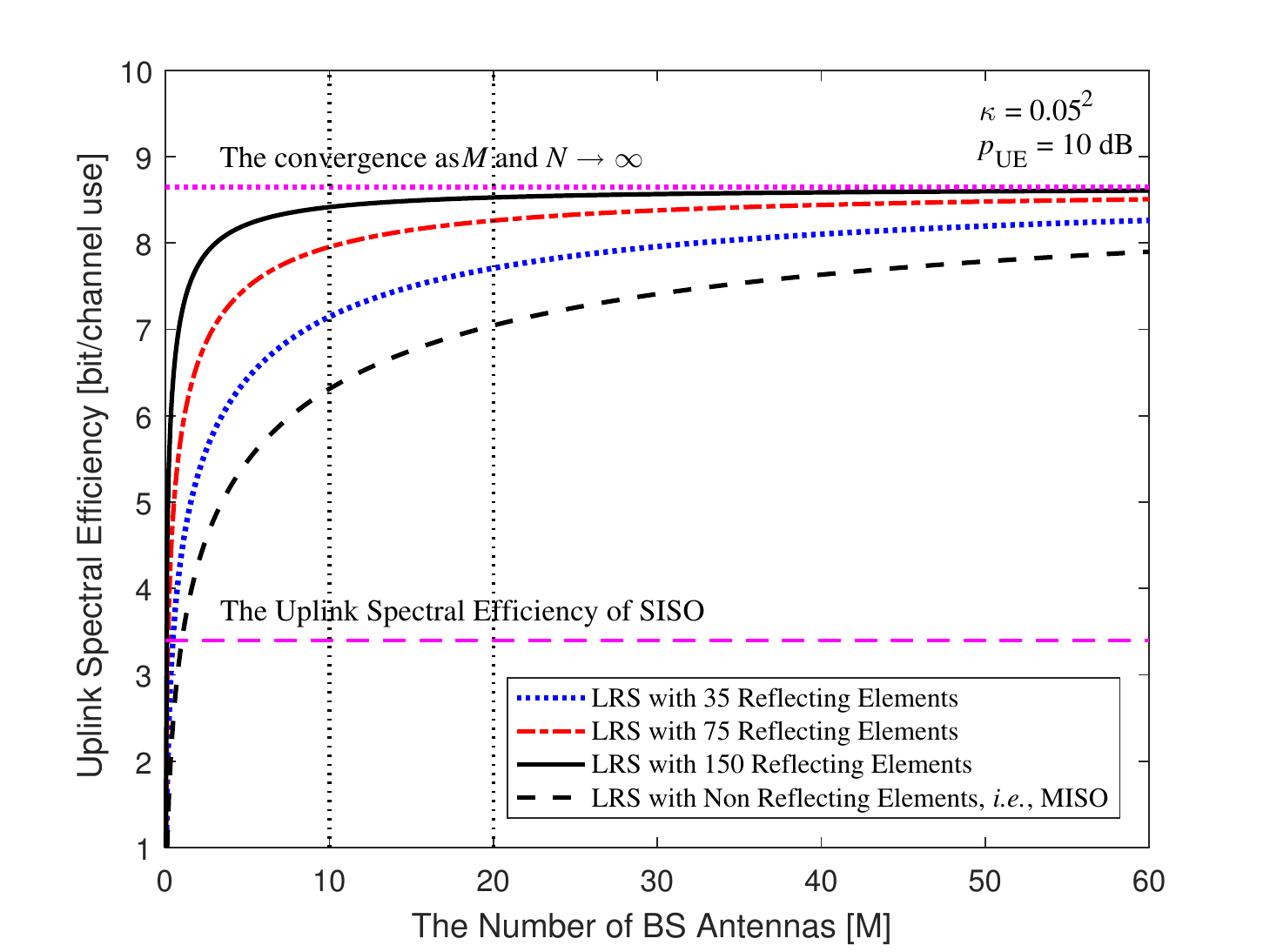}}
	\caption{Spectral efficiency versus the number of BS antennas for $ \kappa_{\mathrm{BS}}=\kappa_{\mathrm{UE}}=0.05^2$ and $ {p_{\mathrm{UE}}}/{\sigma_{\mathrm{BS}}^2} = 10 \text{ dB} $ with different numbers of reflecting elements.}
	\label{fig6}
\end{figure}

To numerically illustrate the power scaling law in LRS-assisted wireless communication system, we compare the spectral efficiency of LRS-assisted system with that of MISO and SISO system. Fig.~\ref{fig5} shows the spectral efficiency on the uplink versus the SNR for $ M=20 $, $ N=100$ and $ \kappa_{\mathrm{BS}}=\kappa_{\mathrm{UE}}=0.05^2$ with perfect and imperfect channel state information, and the SNR is defined as $ {p_{\mathrm{UE}}}/{\sigma_{\mathrm{BS}}^2} $. The LRS-assisted system can reach the limit of spectral efficiency caused by hardware impairments much faster than MISO and SISO system, \textit{i.e.}, have a high spectral efficiency at low SNR. Fig.~\ref{fig6} shows the spectral efficiency versus the BS antennas for $ \kappa_{\mathrm{BS}}=\kappa_{\mathrm{UE}}=0.05^2$ and $ {p_{\mathrm{UE}}}/{\sigma_{\mathrm{BS}}^2} = 10 \text{ dB} $ with different numbers of reflecting elements of LRS. The spectral efficiency increases with the increase of the numbers of BS antennas and reflecting elements, and converges to a finite value given above. These results confirm the fact that we can scale down the transmit power of user as the power scaling laws given in Proposition 1 and Proposition 2.

\section{Conclusion}
In this paper, we study the LRS-assisted communication system by considering hardware impairments. In specific, we study the channel estimation performance as well as the power scaling law in both cases of perfect and imperfect channel state information. The result is encouraging for that we can use more low-cost reflecting elements instead of expensive antennas to achieve higher power scaling. There are other important issues that are not addressed,\textit{ e.g.}, the estimation error increases with the increase of the number of reflecting elements, the estimation error of LRS channel is larger than that of direct channel. These problems cause the demand for more accurate estimation methods and more efficient communication protocols in future works.

\balance
\bibliographystyle{IEEEtran} 
\bibliography{reference}

\end{document}